\documentclass[12pt]{iopart}
\usepackage{color}
\usepackage{graphicx}
\usepackage{float}
\usepackage{subfigure}
\usepackage[normalem]{ulem}
\usepackage{cite}

\usepackage{xcolor}

\begin{document}

\title{Evolution of shot noise in suspended lithographic gold break junctions with  bias and temperature}

\author{Ruoyu Chen$^{1}$,  D.~Natelson$^{1,2,3}$}
\ead{natelson@rice.edu}
\thanks{corresponding author}

\address{$^{1}$ Department of Physics and Astronomy, Rice University, 6100 Main St., Houston, 
TX 77005, USA}

\address{$^{2}$ Department of Electrical and Computer Engineering, Rice University, 6100 Main St.,
Houston, TX 77005, USA}

\address{$^{3}$ Department of Materials Science and NanoEngineering, Rice University, 6100 Main St., Houston, TX 77005, USA}

\begin{abstract}
Shot noise is a powerful tool to probe correlations and microscopic transport details that  conductance measurements alone cannot reveal. Even in atomic-scale Au devices that are well described by Landauer-B{\"u}ttiker physics, complications remain such as local heating and electron-phonon interactions.  We report systematic rf measurements of shot noise in individual atomic-scale gold break junctions at multiple temperatures, with most bias voltages well above the energy of the Au optical phonon mode. Motivated by the previous experimental evidence that electron-phonon interactions can modify Fano factors and result in kinked features in bias dependence of shot noise, we find that the temperature dependence of shot noise from 4.2~K to 100~K is minimal. Enhanced Fano factors near 0.5$~G_0$ and features beyond simply linear bias dependence of shot noise near the 1$~G_0$ plateau are observed. Both are believed to have non-interacting origins and the latter likely results from  slightly bias-dependent transmittance of the dominant quantum channel.

\end{abstract}

\vspace{2pc}

\maketitle

\section{Introduction}
Atomic-scale metal junctions\cite{Agrait:2003} are narrow constrictions with only one or a few metal atoms in cross-section. Such a size is comparable to the metal Fermi wavelength and much smaller than both electron-electron and electron-phonon inelastic mean free paths, even when the constriction itself has already reduced those mean free paths down to the nanometer scale\cite{Erts:2000,Ludoph:1999}.  Thus electron transport through such narrow constrictions is ballistic\cite{Agrait:2003,Beenakker:1991} and is expected to be well described by Landauer-B{\"u}ttiker physics.   This means that (i) electrons transport via distinct quantum channels which, in atomic-scale junctions, originate from valence electronic orbitals\cite{Scheer:1998}, and (ii) inelastic interactions are negligible, such that electrons remain in their own quantum channels\cite{Datta:1995}. The conductance can be expressed as

\begin{equation}
G=G_{0}\sum_{i}^N{\tau_i}
\label{eq:laudauer}
\end{equation}

where $G_0 \equiv2e^2/h$ is the conductance quantum with spin degeneracy, and $\tau_i$ is the transmittance of the $i$th channel. Each channel at most contributes 1~$G_0$ to the total conductance. The success of the Laudauer-B{\"u}ttiker picture does not mean that inelastic interactions can have no observable effect, however. While the constriction itself is small enough to be considered ballistic, electron-phonon interactions can take place within the adjacent contacts, and electrons can potentially couple to vibrational modes localized to the constriction, slightly modifying the total conductance. Well in the tunneling regime ($\tau \ll 0.5$) the electron-phonon interaction permits additional conduction processes and enhances the differential conductance, which is the mechanism of inelastic electron tunneling spectroscopy (IETS)\cite{Jaklevic:1966,Stipe:1998}.  In more conductive single-channel junctions with transmittance $\tau > 0.5$\cite{Paulsson:2005,Tal:2008}
inelastic phonon processes instead reduce the differential conductance, which is used to map out the optical phonon energies in point contact spectroscopy\cite{Agrait:2002,Smit:2002}, when universal conductance fluctuations are believed to be weak.  In both cases, vibrational modification of differential conductance is small and $d^2I/dV^2$ is usually used to detect the tiny changes.  The features in differential conductance occur at the bias voltages $eV$ such that electrons have enough energy to excite one vibrational mode with a certain energy $\hbar\omega$.

Nonequilibrium current fluctuations under bias that originate from the discrete nature of electrons, ``shot noise"\cite{Blanter:2000}, enable access to additional information that conductance alone does not provide.  In its classical limit, as revealed by Schottky in 1918\cite{Schottky:1918}, where electrons transport independently and arrive as a Poisson process, the spectral density of shot noise obeys $S_I=2e \langle I \rangle$.  Here $e$ is the charge of charge carriers and $\langle I \rangle$ is the average bias current.  Correlations alter the shot noise to be $S_I=2e^{\ast} \langle I \rangle F$, with $e^{\ast}$ representing the effective charge and $F$ defined as the Fano factor.  In interacting systems shot noise provides direct information about the effective charge\cite{Saminadayar:1997,dePicciotto:1997,Reznikov:1999,Jehl:2000}; In the Landauer-B{\"u}ttiker limit (where $e^{\ast}=e$) for atomic-scale junctions at zero temperature, the non-interacting shot noise can be fully described by transmittances of quantum channels\cite{Buttiker:1990,Landauer:1991},

\begin{equation}
S_I=2eVG_0\sum_i^N{\tau_i(1-\tau_i)}
\label{eq:shot noise zero T}
\end{equation}

Combining (1) and (2), the corresponding Fano factor is 

\begin{equation}
F=\frac{\sum_i^N{\tau_i(1-\tau_i)}}{\sum_i^N{\tau_i}}
\label{eq:Fano factor}
\end{equation}

At finite temperature where thermal Johnson-Nyquist noise\cite{Johnson:1928,Nyquist:1928} is nonzero, the total current noise including both Johnson-Nyquist and shot noise contributions is

\begin{equation}
S_{I}=G_0 \left[ 4k_{\mathrm{B}}T\sum_i^N{\tau_i^{2}} + 2eV \coth 
\left(\frac{eV}{2k_{\mathrm{B}}T}\right) \sum_{i}^N{\tau_i(1-\tau_i)} \right] .
\label{eq:shot noise finite T}
\end{equation}

Up to moderate biases, the non-interacting predictions of shot noise agree well with experimental observations in break junctions or other mesoscopic systems such as electrostatically defined 2d electron gas (2DEG) structures, and at both cryogenic conditions and room temperature\cite{Reznikov:1995,Kumar:1996,vandenBrom:1999,Djukic:2006,Chen:2012,Vardimon:2013}.  The detailed role of electron-phonon effects in junctions far from equilibrium remains an open question, however.  Multiple theoretical predictions exist\cite{Chen:2005,Schmidt:2009,Avriller:2009,Haupt:2009,Urban:2010,Novotny:2011}, each considering the coupling between electrons of single quantum channel and a single local phonon mode. The main physical difference between the models lies in the basic question of how to properly describe the phonon population of that bosonic mode.  Thermally equilibrated phonon populations, thermally nonequilibrated phonon populations, and nonequilibrated phonon populations (driven into non-thermal steady state by the electronic current) corrected by their fluctuating dynamics are discussed.  The relative importance between electron-phonon coupling strength $\lambda$ and the relaxation rate $\eta$ at which the local vibrational mode loses energy to the bulk phonons in realistic devices determines which regime is experimentally relevant. These theories all predict a modification of the Fano factor when the bias voltage across the junction exceeds the energy of the local vibrational mode. The exact modification of the Fano factor depends on the model details.  Depending on the model, some complicated temperature dependence is predicted.  Kumar \textit{et al.}\cite{Kumar:2012} observed kink-like features in the bias dependence of shot noise in many atomic-scale Au junctions at 4.2~K. These kinks are usually observed at a voltage between 10 to 20~mV, consistent with the known energy of the optical phonon in Au\cite{Frederiksen:2004,Agrait:2002} and the measured differential bias-dependent changes in the conductance. If the inelastic contribution to the noise is detectable, a temperature dependent study is favored, as the phonon population plays an important role.  Such a study has not previously been performed in atomic-scale devices.

In previous work we have examined shot noise in STM-style Au break junctions at room temperature, where $k_{\mathrm{B}}T$ readily exceeds the metal optical phonon energy\cite{Chen:2012}.  In this regime, the bias dependence of the \textit{ensemble-averaged} noise is consistent with Landauer-B{\"u}ttiker expectations at biases below about 200 mV.  Since weak electron-phonon interactions should be small, and the coupling strength $\lambda$ between electrons and the vibrational modes depends on the microscopic junction details, ensemble averages might wash out modest modifications to the noise.  Measurements of individual electromigrated junctions at reduced temperature have shown nonlinearities and asymmetries in the noise at 77~K.  Detailed measurements over many individual stable microscopic configurations are needed to see whether temperature strongly affects the noise bias dependence in the moderate bias regime\cite{Wheeler:2013}. Here, we report bias dependences of shot noise in e-beam lithographically patterned, suspended Au junctions, over large bias and conductance ranges at multiple temperatures, realized by the mechanically controlled break junction (MCBJ) technique\cite{Muller2:1992,Muller:1992}. We find that temperature does not strongly influence the noise bias dependence in these junctions beyond what is expected from thermal broadening of the electronic distributions.  These noise measurements also allow us to confirm that the transmission and mixing of channels during junction rupture is largely unchanged over a broad temperature range.

\section{Methods}
Stainless steel is chosen to be the substrate due to its flexibility, with a polyimide (PI-2610) film coated as the insulating layer.  Arrays of bowtie junctions are defined by e-beam lithography and e-beam evaporation (Ti/Au). The narrowest part of a typical bowtie junction is a Au bridge of $\sim$ 800~nm long, $\sim$150~nm wide and 20~nm thick, which typically has a resistance $\sim 70~\Omega$ at room temperature.  After the lift-off, thick Au pads and leads connecting the junctions are evaporated making use of a shadow mask with Cr as the adhesion layer.  Ar plasma treatment is needed before the second evaporation to improve adhesion. Poor adhesion between metal pads and polyimide has previously made wire bonding normally impossible. The combination of Ar plasma treatment, adhesion layer material selection of Cr, and comparatively thick bonding pads together make wire bonding reliable.  Before the measurement, O$_2$ plasma etching is used to suspend the narrow Au bridges, leaving the other metal regions slightly undercut. The SEM photo of a typical suspended junction is shown in figure 1(a).

Measurements are conducted in our home-built variable temperature setup, which can be cooled down to liquid He temperature via gas flow and has a stepper motor configured to bend the chip. The chip is immersed in cold He gas with positive pressure, and temperature-controlled with a local heater.  The Au bridges are first narrowed down by electromigration, and then further broken by slowly bending the substrate, using the conventional MCBJ technique.  We stop at particular conductance values and record the bias current and the excess noise simultaneously as a function of bias voltage. Reconnections to $\sim$ $10~G_0$ are needed to effectively randomize the microscopic structures. The small size of the suspended part of the device and the mechanical bending geometry leads to greatly enhanced mechanical stability of the Au junctions when compared to STM-style junctions.  This stability is often characterized by a ``reduction factor'', $r$, the ratio between the lateral displacement of the electrode tips and the vertical displacement of the center of the bending setup\cite{Vrouwe:2005}.  In its simplest treatment, $r = 6tU/L^{2}$, where $t$ is the substrate thickness (including the polyimide), $U$ is the length of the suspended region of junction, and $L$ is the distance between the fixed support posts for the bending scheme.  For our situation, $t \approx 100\mu$m, $U \approx$~800~nm, and $L \approx$~1.5~cm, leading to $r \approx 2 \times 10^{-6}$.

All the measurements are conducted on fixed stable atomic configurations, avoiding any transients associated with changes in junction geometry during a particular noise measurement. The RF modulated excess noise measurement technique has been described previously\cite{Reznikov:1995,Wu:2007,Wheeler:2010,Chen:2012}.  We apply 1~kHz square waves switching between zero and a finite voltage to the device and the bias-driven change in rf noise power (integrated over a bandwidth set by filters to be from 250 to 500 MHz) is measured via lock-in amplifier. The DC signal is used to measure the conductance. Shot noise should dominate in the bandwidth measured. The conductive substrate and the thin layer of polyimide turn out to be an effective capacitor and contribute to RF signal loss in our bandwidth, which limits our measurement resolution in this work, but could be potentially improved in the future.  To calibrate the noise to obtain the appropriate normalization, we consider measurements in the tunneling region with conductance smaller than $0.1~G_0$, where the Fano factor is $1-\tau$ and $\tau$ is the only relevant quantum channel.  The damping of the measured rf noise due to the substrate is roughly consistent with the measured rf reflectance of the device, though detailed quantitative modeling has proven difficult.

\section{Results and discussion}
At zero temperature, shot noise is linear in bias, if the Fano factor is independent of bias. At finite temperatures, in the non-interacting limit, low bias ($eV\ll2k_{\mathrm{B}}T$) excess noise ($S_I(V)-S_I(V=0)$) is suppressed while the higher bias excess noise still scales linearly with the bias, as shown in figure 1(b). The excess noise remains linearly proportional to a scaled bias $X\equiv 4k_{\mathrm{B}}TG[(eV/2k_{\mathrm{B}}T)\textup{coth}(eV/2k_{\mathrm{B}}T)-1]$, with the slope exactly given by the transmittance-based Fano factor. In the presence of electron-phonon interactions, the Fano factor is expected to exhibit a bias dependence above the electron-phonon onset voltage, $V= \hbar \omega/e$.  Theories considering the coupling between electrons and thermally nonequilibrated vibrational modes (electron-phonon scattering rate $\Gamma_{e-ph}\gg\eta$)\cite{Haupt:2009,Urban:2010,Novotny:2011} all predict that the bias dependence has components scale as $V^2$ or faster. Theories considering only thermally equilibrated vibrational populations ($\Gamma_{e-ph}\ll\eta$)\cite{Avriller:2009, Haupt:2009,Schmidt:2009} instead predict an abrupt change of $F$ at electron-phonon interaction onset voltage and slower V dependence. Deviations from linearity in plots of $S_{I}(V)-S_{I}(0)$ vs. $X$ can be indicative of these inelastic processes.

Measurements are conducted with sample temperatures between 4.2~K to 100~K.  As the theories generally discuss single channel transport, and additional quantum channels complicate even the non-interacting interpretation, we mostly focus on junction configurations with conductances below $2~G_0$.  Because the shot noise signal scales with current, the noise is most clearly resolved at bias voltages that are well in excess of $\sim17$~mV, the expected threshold for the Au optical phonon. Thus the deduced Fano factors should already be modified by inelastic effects, if any are present. We restrict applied biases to less than 200 mV to maintain junction stability; hence the nonlinear increase in noise at higher biases\cite{Chen:2014} is not examined here.

We performed detailed measurements in different conductance regimes: the tunneling regime with $G<0.1~G_0$, the single-atom contact regime with $G\sim 1~G_0$, and the few-channel point contact regime, $1~G_0<G<10~G_0$.  Au junction stability does not favor the conductance values between $0.1~G_0$ and much below $1~G_0$, though repeatedly disconnecting and reconnecting the contacts sometimes results in configurations with sufficient stability for data collection.  The scaled bias dependence of the noise found in most junction configurations is linear with approximately vanishing intercepts. A typical example is shown in figure 1(c), where the environmental temperature is $\sim$45~K. The exact value of the temperature matters in the scaled-bias fitting, as shown in the inset of figure 1(c); only the correct temperature minimizes the intercept in a linear fit, which is also the main basis of the shot noise thermometry\cite{Spietz:2003}.  Consistency between the measured sample temperature (via conventional thermometry) and the temperature parameter that linearizes (with zero intercept) noise vs. scaled bias indicates the high data quality and the dominance of shot noise. 
\begin{figure}[htb] 

\includegraphics[width=1\textwidth]{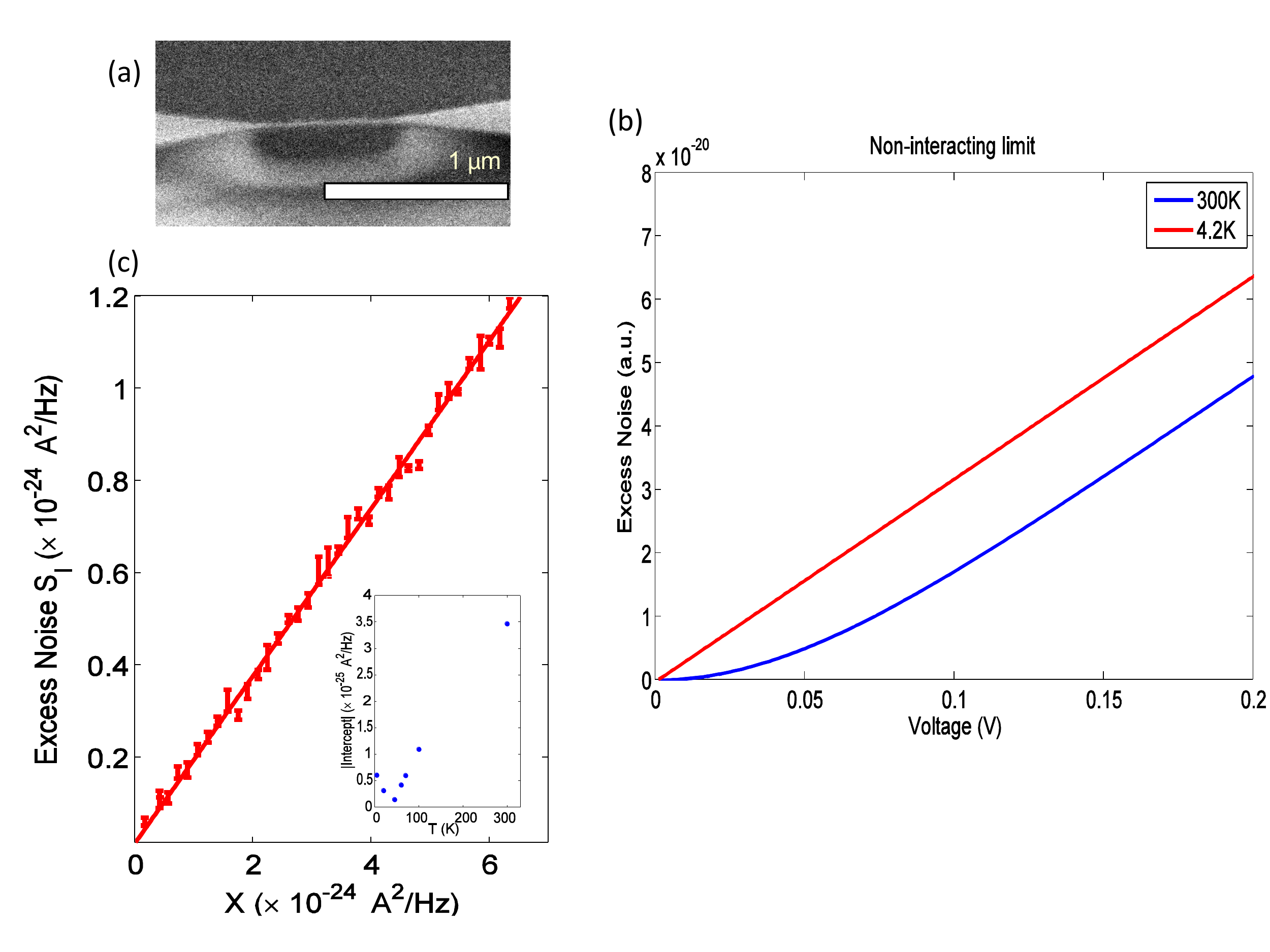}                                                             

\caption{ (a) An electron micrograph of a suspended bowtie junction. (b) The calculated non-interacting excess noise at 4.2~K and 300~K, with Fano factor a constant. The same Fano factor yields the same high bias slopes, while the elevated environmental temperature suppresses the low bias excess noise spectral density. (c) An example of linear scaled-bias dependence, with the red line the best linear fitting result. This example data is of a 2.4$~G_0$ device, taken at 45~K. In the bottom inset, if the wrong temperature is used in the scaled bias, the linear fit will generate a larger intercept (and the low bias data will deviate visibly from the fit). It's clear that the minimized intercept is acheived at $\sim$45~K. This panel is a self-consistency check of the validity of the linear fitting analysis. 
}
\label{Fig.1}
\end{figure}

The lack of clear nonlinearities means that no significant effects from thermally nonequilibrated vibrational modes are observed in most devices. Previous experiments did suggest the local heating of phonon might be weak in Au atomic-scale junctions\cite{Chen:2014,Adak:2015}. In a small fraction of the measurements we do see small upwards nonlinearities close to the highest applied voltages, at all temperatures. This could be a result of unusually strong coupling with nonequilibrated vibrational modes, weak energy relaxation associated with some particular atomic configurations, or junction instability driven by bias.  Given the challenge of interpreting these idiosyncratic configurations, we focus on the configurations that do not show these nonlinearities. 

In typical, stable junctions, the Fano factors could be extracted by the linear fitting procedure, allowing us to generate a map of the distribution of Fano factors, shown in figure 2(a). Close to the $1~G_0$ plateau, some recorded bias dependences have small nonlinearities that result in uncertainties when determining the Fano factors, but will not cause significant variations in the exact values.  These features will be discussed in the last section of this paper and are believed to have non-interacting origins.

\begin{figure}[htb]                                                                     

\includegraphics[width=1\textwidth]{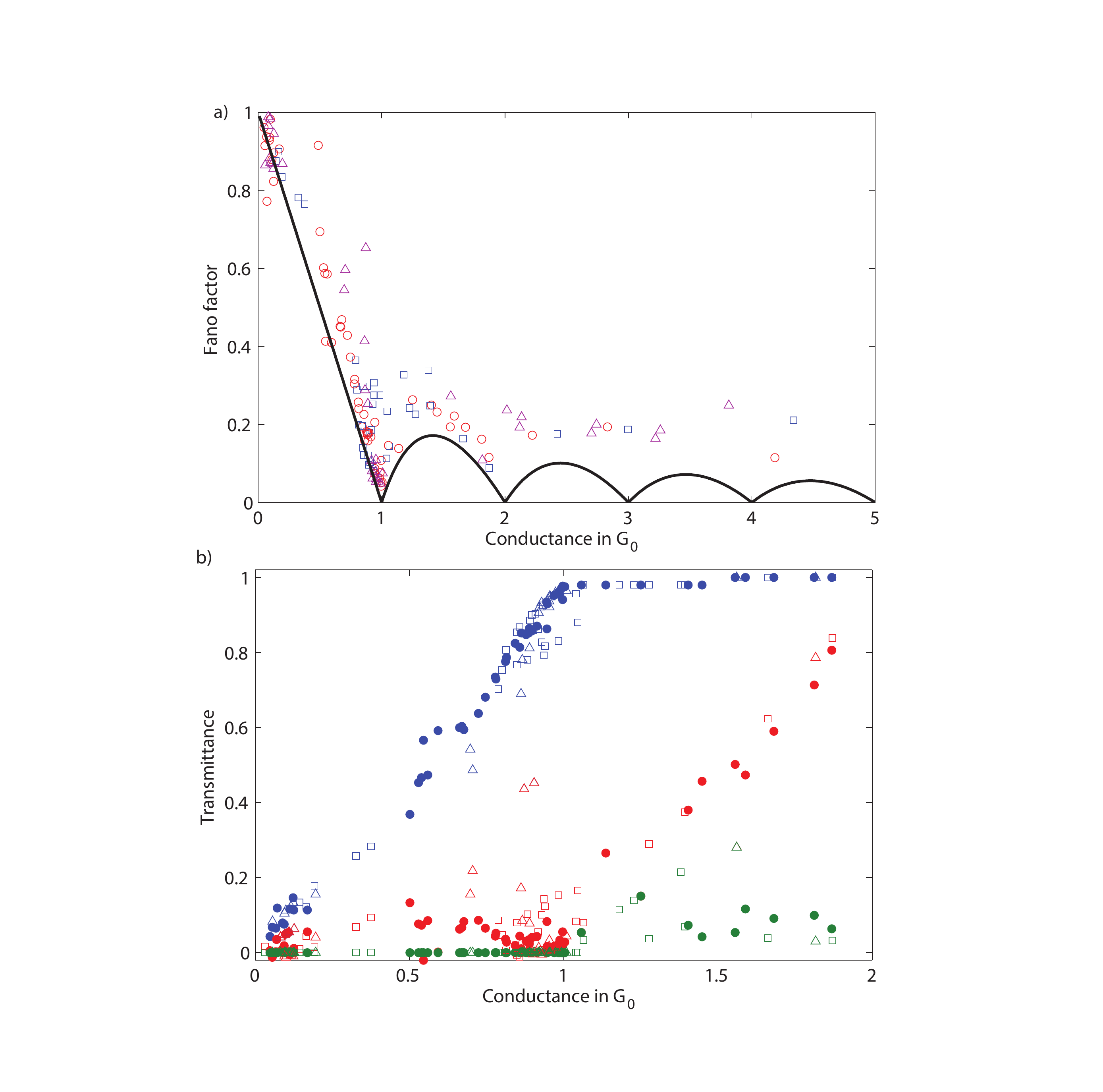}
\caption{ (a) The conductances and Fano factors measured on all the devices. The red circles represent measurements at 4.2~K, and the blue squares and purple triangles represent data at $\sim$ 50~K and 100~K, respectively. The black curve is the non-interacting ``forbidden region'' boundary. (b) The deduced transmittances of the first three quantum channels from panel (a). Blue, red, green colors represent $\tau_1$, $\tau_2$ and $\tau_3$. Filled circles represent data at 4.2~K, and the open squares and triangles represent the data at $\sim$ 50~K and 100~K, respectively. }
\label{Fig.2}
\end{figure}

In figure 2(a) red circles represent measurements at 4.2~K, while blue squares and purple triangles are the data at 50~K and 100~K, respectively. The black curve is the mathematically allowed smallest shot noise in non-interacting limit, reached when only one quantum channel has transmittance not equal to 1. Atomic-scale Au junctions are well approximated by this limit, especially when $G < 1~G_0$.  The region below the black curve is forbidden in the Landauer-B{\"u}ttiker model with spin degeneracy\cite{Kumar:2013,Vardimon:2015}. Beyond experimental uncertainty, no measured points fall in this region. Most data are distributed close to the forbidden region boundary and two pronounced quantum suppressions of Fano factors are seen, at $1~G_0$ and close to $2~G_0$.  All these observations are consistent with previous experiments that show that clean Au atomic-scale junctions favor fully transmitted channels. Interestingly, the distribution of Fano factors does not seem to depend significantly on temperature.   Especially when $G\rightarrow 0,~1~G_0$, the measured data at different temperatures overlap with each other very well. While the 4.2~K data seems follow the same trace and have very small variation, a few outliers at higher temperatures do have higher Fano factors. This likely results from channel mixture and greater likelihood of junction contamination by adsorbates at higher temperature.

Over the temperature range 4.2~K to 100~K, the corresponding vibrational occupations $n_B=1/(e^{\beta\hbar\omega}-1)$ increase from $3\times10^{-21}$ (essentially zero) to 0.16, if taking $\hbar\omega=0.17~\textup{meV}$. Even as small as an occupancy of 0.16 might generate an additional noise larger or comparable to the total inelastic correction at zero temperature limit\cite{Haupt:2009,Urban:2010} and should be detectable. The linear bias dependence in our measurements already indicates little inelastic contribution to the noise. The lack of phonon population dependence further indicates apparently weak e-ph coupling in our devices. To further clarify the effect of thermal phonon occupation measurements at higher temperatures might be favored, but the stability of atomic-sized devices naturally limits practical environmental temperatures and make such measurements challenging.

Regardless of temperature dependence, theories considering the interactions between electrons of single quantum channel and single thermally equilibrated vibrational mode predict that Fano factors are suppressed when the single channel transmittance is between $1/2\pm1/2\sqrt{2}$ and enhanced otherwise\cite{Schmidt:2009,Avriller:2009}, even at zero temperature. This is because electrons can always emit phonons regardless phonon population and the only requirement is to have enough energy. Thus in the predicted suppression region, if electron-phonon interactions cause a significant modification, we might have a chance to see some data fall below the non-interacting forbidden region, but this is not seen. Instead, compared with the data with $G\rightarrow 0,1~G_0$, the measured Fano factors in between $1/2\pm1/2\sqrt{2}~G_0$ are enhanced relative to the black curve. Such a positive deviation is likely to be a result of non-interacting channel mixture, which potentially hides any effects from electron-phonon interactions. One possible origin of such a channel mixture is adsorbate contamination. Indeed, geometrically Au devices do not favor conductance values between 0 and $1~G_0$. The reason stable configurations in this conductance region sometimes still can be reached might be a result of contamination, which itself could also induce channel mixture.

Assuming the noise results are well described by the Landauer-B{\"u}ttiker model, combining the conductance and Fano factor measurements we are able to calculate the transmittances of the first three quantum channels below $2~G_0$, as what we show in figure 2(b). We assume the third channel is not opened below $1~G_0$ ($\tau_3$=0), and the first channel is always nearly fully transmitted above $1~G_0 $ ($\tau_1\rightarrow1$). Thus at all the conductance values below $2~G_0$ we always only have two variables and independent measurements of noise and conductance determine them completely. When $1~G_0<G<2~G_0$ we have to manually set the first channel's transmittance $\tau_1$ to be a value slightly lower than 1 (but larger than 0.95), to force the calculated $\tau_2$ and $\tau_3$ to be real. This introduces some uncertainty but will be qualitatively valid. This implies that the first quantum channel is not fully opened near $1~G_0$ in our devices, which is consistent with previous observations not showing perfect noise suppression\cite{Armstrong:2010,Vardimon:2013,Chen2:2014}. As a self-consistency check, $\tau_2$ and $\tau_3$ are indeed small below $1~G_0$ and $2~G_0$ respectively. Simulations in a previous paper\cite{Chen2:2014} based on molecular dynamics and the Green functions technique match these observations. At 4~K such a simulation predicted an enhancement of the ensemble-averaged Fano factors close to $0.5~G_0$, as a result of maximized $\tau_2$ contribution; This is reproduced by the transmittance mapping extracted from the data. This good agreement favors the channel mixing interpretation and confirms that the intrinsic motions of Au atoms could also be a possible origin of the enhanced Fano factors between 0 and $1~G_0$, as well as the non-saturated quantum channel near $1~G_0$. On the other hand, the same simulation suggests that the second channel's contribution below $1~G_0$ will be largely washed out at room temperature, dut to changes in configuration stability. Such a difference is not seen experimentally up to 100~K.

We also find that the bias dependence of shot noise close to the $1~G_0$ plateau is sometimes slightly curved and seems to have reproducible deviations from linearity as a function of bias. Two examples are shown in figure 3(a-d) and 3(e-h). Data points in both bottom panels 3(d) and 3(h) with error bars represent the measured excess noise spectral density as a function of bias. Top panels 3(a) and 3(e) are the simultaneously measured I-V curves. The example shown in left column 3(a-d) is a $\sim0.89~G_0$ device with the second channel participation $\tau_2$ as small as 0.03. The measured noise clearly is concave upward. The right column 3(e-h) represents another device at $\sim0.94~G_0$ but has a significant contribution (0.14) from $\tau_2$ and less nonlinearity with bias for both $G$ and $S_I$.

At each bias, the measured I-V curves, though still pretty linear, have small but highly reproducible (from sweep to sweep) deviations from linearity. Such small deviations are easier to see in panels 3(b) and (f), where the difference between the measured conductance data and the linear fittings (shown as solid straight lines in 3(a) and (e)) are plotted as a function of bias. This indicates a weak bias dependence of the conductance, or the transmittances $\{\tau_i\}$ of the single-atom constriction, as what has been shown in 3(c) and (g). As a result, based on the equation (\ref{eq:Fano factor}) Fano factors can vary as a function of bias and even the non-interacting shot noise can have more complicated nonlinear bias dependence, which could be easily calculated.  Indeed when larger variations of conductance over bias are observed, the junctions tend to have a more curved dependence of noise with scaled bias, as shown in 3(a-d). In junctions whose conductance have smaller variations, like the device in 3(e-h), the scaled bias dependence of the noise tends to be linear and featureless.

Universal conductance fluctuations (UCF) provide a mechanism to explain such a weak bias dependence of the $\{\tau_i\}$. Though below 0.9~V there won't be significant variations of the transmittances in single-atom Au constrictions\cite{Nielsen:2002} and the density of states of Au in the relevant energy range are relatively featureless,disorder and quantum interference can affect the $\{\tau_i\}$\cite{Ludoph:1999}.

\begin{figure}[htb] 

\includegraphics[width=1\textwidth]{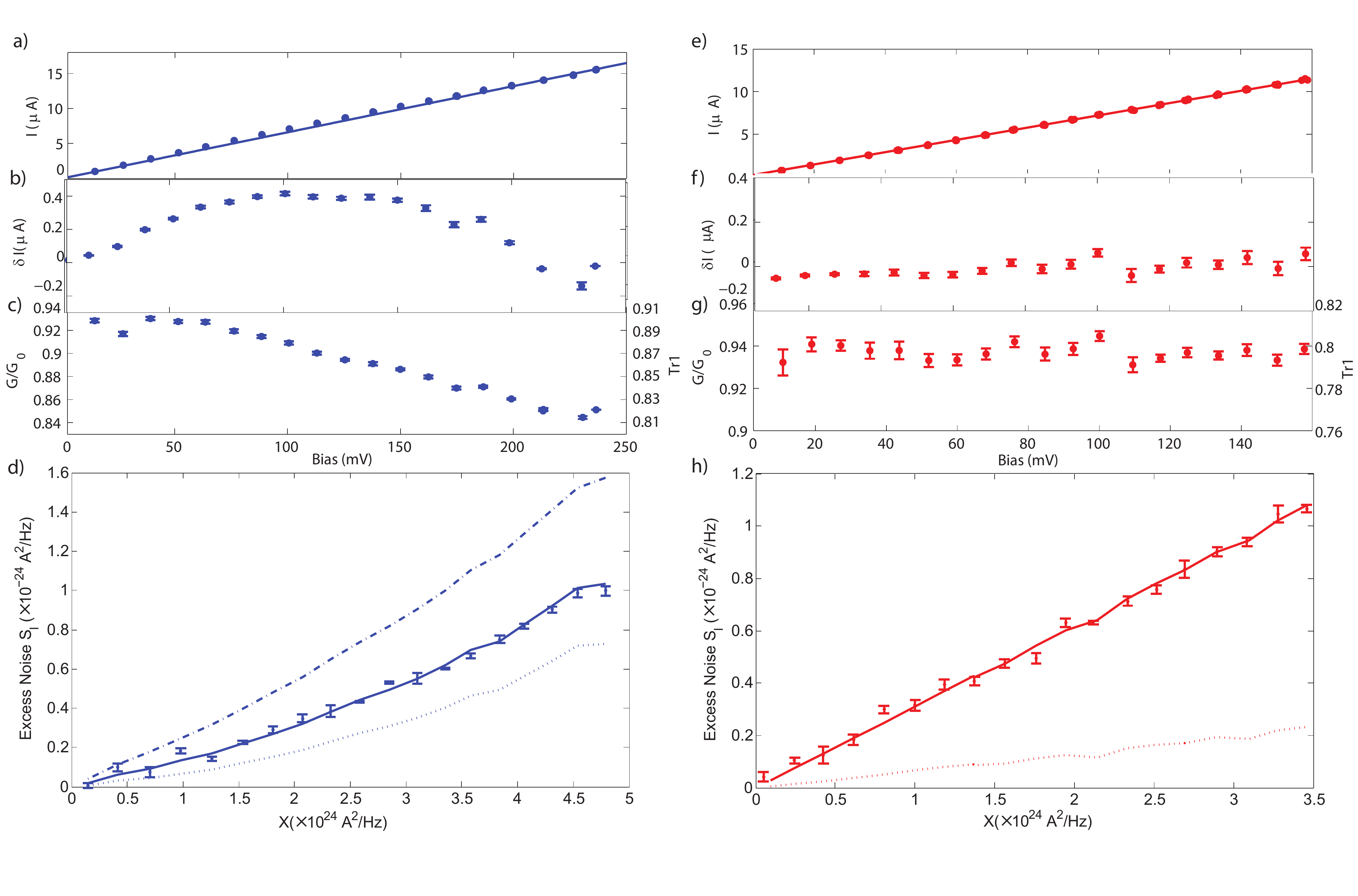}                                                             

\caption{Two examples of the nonlinear bias dependence near $1~G_0$: (a-d) $\sim$ 0.89~$G_0$ with the best-fitted $\tau_2\sim0.03$. The conductance has a relatively strong bias dependence.  (e-h) $\sim$ 0.94~$G_0$ with the best-fitted $\tau_2\sim0.14$. The conductance has weaker bias dependence.  From top to bottom, (a)(e) I-V curve, (b)(f) deviation $\delta I\equiv~I-I_{linearity}$ vs V, (c)(g) G(V) vs V and (d)(h) the measured excess noise vs the scaled bias respectively.  In the bottom panel (d)(h) the solid curve represents the expected shot noise after properly accounting for the bias dependence of device conductance and choosing $\tau_2$ to be the best-fitted value. The dotted line forces $\tau_2$ to be zero (single channel case), while the dashed curve in (d) forces $\tau_2=0.1$. Based on the best-fitted value of $\tau_2$, the $\tau_1$ values associated to the total conductance are also indicated in panel (c) and (g).}
\label{Fig.3}
\end{figure}

To verify the correlation between the observed nonlinear features in shot noise and the conductance fluctuation, we recalculate the expected shot noise, allowing the conductance to be weakly bias dependent. The single-atom Au constriction can be well approximated by a single quantum channel picture, and a more complete description allows the second channel to be weakly transmitted. Thus we have $G/G_0=\tau_1+\tau_2$, with $G\rightarrow 1~G_0$, $\tau_1 \gg \tau_2$. To model the slightly bias dependent conductance, we have G=G(V)=$<G>+\delta G(V)$. The ideal procedure to extract G(V) is to densely measure the differential conductance as a function of bias, and integrate the result from zero to the target voltage. Our apparatus did not allow simultaneous noise and differential conductance measurements; instead, we approximated this by taking $\Delta I/\Delta V$ at each bias from the $I-V$ curve, though this sacrifices the voltage resolution.

By plugging $G(V)$ into equation (\ref{eq:shot noise finite T}), the expected shot noise spectral density $S_I=S_I(\tau_2, X)$ was calculated. A least square fitting to the data generates the best-fitted value of $\tau_2$ and the $S_I(\tau_2, X)$ is reduced to $S_I(X)$. The best-fitted results of both examples are plotted as the solid curves in figure 3(d) and (h), retracing the data reasonably well. This simple toy model calculation can be used to describe varying degrees of nonlinearity in $G(V)$, and was successfully applied to all other devices not shown here. Consistency between the independent measured $G(V)$ and $S_I(V)$ when analyzed this way strongly suggests that such a voltage dependent $\{\tau_i\}$ can explain both data sets. A similar analysis has been performed by others (unpublished, Vardimon and Tal \textit{et al.} \cite{Vardimon2:2015}).

It is striking that such small nonlinearities in $I$ vs $V$ can result in a readily apparent effect on shot noise, though this is mathematically natural. The shot noise carried by each individual quantum channel is proportional to $\tau_i(1-\tau_i)$. Considering a small variation $\delta \tau_i$, the resulting shot noise is $(\tau_i+\delta\tau_i)(1-\tau_i-\delta\tau_i)$. As the transmittance of the dominant channel $\tau_1\rightarrow 1$, $\tau_1+\delta \tau_1$ is almost unchanged but $\delta \tau_1$ can be comparable to $1-\tau_1$ and leads to a change by a significant fraction. If $\tau_1\rightarrow 1$ is not valid, shot noise becomes comparatively insensitive to the small $\delta \tau_1$ and the normal linear bias dependence of $S_I$ is recovered, as what we have seen away from the $1~G_0$ plateau. Similarly, if $\tau_2$ is non-negligible, even at the $1~G_0$ plateau $\tau_1$ (due to the fact that $\tau_1=G/G_0-\tau_2$) is not close to 1 and shot noise tends to be less sensitive to the conductance fluctuations.

\section{Conclusion}

In summary, we examined the bias dependence of shot noise at atomic-scale Au junctions, over a large bias, conductance, and temperature range. Even at the biases well above the energy of the Au optical phonons and varying the temperature from 4.2~K to 100~K, no clear temperature dependence is seen. This likely indicates weak inelastic contribution to the shot noise associated with very small electron-phonon coupling strength $\lambda$ and thermal phonon populations. Fano factors are observed to be enhanced near $0.5~G_0$, which either results from contamination or the intrinsic atomic configurations favored by Au junction formation. Near the $1~G_0$ plateau, the bias dependence of shot noise shows geometry-dependent small nonlinearities, which are well described by a simple non-interacting scenario considering weak bias dependence of transmittances as expected from UCF.

\section{Acknowledgments}

D.N.\ and R.C.\ acknowledge support of NSF awards DMR-1305879, as well as the helpful discussions with Ran Vardimon. We also acknowledge the earlier work of Leland Richardson for the design and construction of the variable temperature measurement system.

\clearpage


\begin{thebibliography}{10}

\bibitem{Agrait:2003}
Nicol{\'a}s Agra{\"i}t, Alfredo~Levy Yeyati, and Jan~M. van Ruitenbeek.
\newblock Quantum properties of atomic-sized conductors.
\newblock {\em Physics Reports}, 377(2–3):81 -- 279, 2003.

\bibitem{Erts:2000}
D.~Erts, H.~Olin, L.~Ryen, E.~Olsson, and A.~Th\"ol\'en.
\newblock Maxwell and sharvin conductance in gold point contacts investigated
  using tem-stm.
\newblock {\em Phys. Rev. B}, 61:12725--12727, May 2000.

\bibitem{Ludoph:1999}
B~Ludoph, M.~H. Devoret, D~Esteve, C~Urbina, and J.~M. van Ruitenbeek.
\newblock Evidence for saturation of channel transmission from conductance
  fluctuations in atomic-size point contacts.
\newblock {\em Phys. Rev. Lett.}, 82(7):1530--1533, 1999.

\bibitem{Beenakker:1991}
CWJ Beenakker and Henk van Houten.
\newblock Quantum transport in semiconductor nanostructures.
\newblock {\em Solid state physics}, 44(1):228, 1991.

\bibitem{Scheer:1998}
E.~Scheer, N.~Agra{\"i}t, J.~C. Cuevas, A.~Levy Yeyati, B.~Ludoph,
  A.~Mart{'i}n-Rodero, G.~R. Bollinger, J.~M. van Ruitenbeek, and C.~Urbina.
\newblock The signature of chemical valence in the electrical conduction
  through a single-atom contact.
\newblock {\em Nature}, 394:154--157, 1998.

\bibitem{Datta:1995}
S.~Datta.
\newblock {\em Electronic Transport in Mesoscopic Systems}.
\newblock Cambridge University, Cambridge, UK, 1995.

\bibitem{Jaklevic:1966}
R.~C. Jaklevic and J.~Lambe.
\newblock Molecular vibration spectra by electron tunneling.
\newblock {\em Phys. Rev. Lett.}, 17:1139--1140, Nov 1966.

\bibitem{Stipe:1998}
BC~Stipe, MA~Rezaei, and W~Ho.
\newblock Single-molecule vibrational spectroscopy and microscopy.
\newblock {\em Science}, 280(5370):1732--1735, 1998.

\bibitem{Paulsson:2005}
Magnus Paulsson, Thomas Frederiksen, and Mads Brandbyge.
\newblock Modeling inelastic phonon scattering in atomic-and molecular-wire
  junctions.
\newblock {\em Phys. Rev. B}, 72(20):201101, 2005.

\bibitem{Tal:2008}
O~Tal, M~Krieger, B~Leerink, and JM~van Ruitenbeek.
\newblock Electron-vibration interaction in single-molecule junctions: From
  contact to tunneling regimes.
\newblock {\em Phys. Rev. Lett.}, 100(19):196804, 2008.

\bibitem{Agrait:2002}
Nicol\'as Agra{\"i}t, Carlos Untiedt, Gabino Rubio-Bollinger, and Sebasti\'an
  Vieira.
\newblock Onset of energy dissipation in ballistic atomic wires.
\newblock {\em Phys. Rev. Lett.}, 88:216803, May 2002.

\bibitem{Smit:2002}
RHM Smit, Y~Noat, C~Untiedt, ND~Lang, MC~Van~Hemert, and JM~Van~Ruitenbeek.
\newblock Measurement of the conductance of a hydrogen molecule.
\newblock {\em Nature}, 419(6910):906--909, 2002.

\bibitem{Blanter:2000}
Ya.M. Blanter and M.~B{\"u}ttiker.
\newblock Shot noise in mesoscopic conductors.
\newblock {\em Physics Reports}, 336(1–2):1 -- 166, 2000.

\bibitem{Schottky:1918}
W~Schottky.
\newblock Regarding spontaneous current fluctuation in different electricity
  conductors.
\newblock {\em Ann. der Physik}, 57(23):541--567, 1918.

\bibitem{Saminadayar:1997}
L.~Saminadayar, D.~C. Glattli, Y.~Jin, and B.~Etienne.
\newblock Observation of the $\mathit{e}\mathit{/}3$ fractionally charged
  laughlin quasiparticle.
\newblock {\em Phys. Rev. Lett.}, 79:2526--2529, Sep 1997.

\bibitem{dePicciotto:1997}
R.~de~Picciotto, M.~Reznikov, M.~Heiblum, V.~Umansky, G.~Bunin, and D.~Mahalu.
\newblock Direct observation of a fractional charge.
\newblock {\em Nature}, 389:162--164, 1997.

\bibitem{Reznikov:1999}
M~Reznikov, R~De~Picciotto, TG~Griffiths, M~Heiblum, and V~Umansky.
\newblock Observation of quasiparticles with one-fifth of an electron's charge.
\newblock {\em Nature}, 399(6733):238--241, 1999.

\bibitem{Jehl:2000}
X~Jehl, M~Sanquer, R~Calemczuk, and D~Mailly.
\newblock Detection of doubled shot noise in short normal-metal/superconductor
  junctions.
\newblock {\em Nature}, 405(6782):50--53, 2000.

\bibitem{Buttiker:1990}
M~B{\"u}ttiker.
\newblock Scattering theory of thermal and excess noise in open conductors.
\newblock {\em Phys. Rev. Lett.}, 65(23):2901, 1990.

\bibitem{Landauer:1991}
R.~Landauer and Th. Martin.
\newblock Equilibrium and shot noise in mesoscopic systems.
\newblock {\em Physica B: Condensed Matter}, 175(1–3):167 -- 177, 1991.

\bibitem{Johnson:1928}
J.~B. Johnson.
\newblock Thermal agitation of electricity in conductors.
\newblock {\em Phys. Rev.}, 32:97--109, Jul 1928.

\bibitem{Nyquist:1928}
H.~Nyquist.
\newblock Thermal agitation of electric charge in conductors.
\newblock {\em Phys. Rev.}, 32:110--113, Jul 1928.

\bibitem{Reznikov:1995}
M.~Reznikov, M.~Heiblum, Hadas Shtrikman, and D.~Mahalu.
\newblock Temporal correlation of electrons: Suppression of shot noise in a
  ballistic quantum point contact.
\newblock {\em Phys. Rev. Lett.}, 75:3340--3343, Oct 1995.

\bibitem{Kumar:1996}
A.~Kumar, L.~Saminadayar, D.~C. Glattli, Y.~Jin, and B.~Etienne.
\newblock Experimental test of the quantum shot noise reduction theory.
\newblock {\em Phys. Rev. Lett.}, 76:2778--2781, Apr 1996.

\bibitem{vandenBrom:1999}
H.~E. van~den Brom and J.~M. van Ruitenbeek.
\newblock Quantum suppression of shot noise in atom-size metallic contacts.
\newblock {\em Phys. Rev. Lett.}, 82:1526--1529, Feb 1999.

\bibitem{Djukic:2006}
D.~Djukic and J.~M. van Ruitenbeek.
\newblock Shot noise measurements on a single molecule.
\newblock {\em Nano Lett.}, 6(4):789--793, 2006.

\bibitem{Chen:2012}
Ruoyu Chen, Patrick~J. Wheeler, and D.~Natelson.
\newblock Excess noise in stm-style break junctions at room temperature.
\newblock {\em Phys. Rev. B}, 85:235455, Jun 2012.

\bibitem{Vardimon:2013}
Ran Vardimon, Marina Klionsky, and Oren Tal.
\newblock Experimental determination of conduction channels in atomic-scale
  conductors based on shot noise measurements.
\newblock {\em Phys. Rev. B}, 88:161404, Oct 2013.

\bibitem{Chen:2005}
Yu-Chang Chen and Massimiliano Di~Ventra.
\newblock Effect of electron-phonon scattering on shot noise in nanoscale
  junctions.
\newblock {\em Phys. Rev. Lett.}, 95:166802, Oct 2005.

\bibitem{Schmidt:2009}
TL~Schmidt and A~Komnik.
\newblock Charge transfer statistics of a molecular quantum dot with a
  vibrational degree of freedom.
\newblock {\em Phys. Rev. B}, 80(4):041307, 2009.

\bibitem{Avriller:2009}
R.~Avriller and A.~Levy~Yeyati.
\newblock Electron-phonon interaction and full counting statistics in molecular
  junctions.
\newblock {\em Phys. Rev. B}, 80:041309, Jul 2009.

\bibitem{Haupt:2009}
Federica Haupt, T.~Novotn\'y, and Wolfgang Belzig.
\newblock Phonon-assisted current noise in molecular junctions.
\newblock {\em Phys. Rev. Lett.}, 103:136601, Sep 2009.

\bibitem{Urban:2010}
DF~Urban, R~Avriller, and A~Levy Yeyati.
\newblock Nonlinear effects of phonon fluctuations on transport through
  nanoscale junctions.
\newblock {\em Phys. Rev. B}, 82(12):121414, 2010.

\bibitem{Novotny:2011}
T.~Novotn\'y, Federica Haupt, and Wolfgang Belzig.
\newblock Nonequilibrium phonon backaction on the current noise in atomic-sized
  junctions.
\newblock {\em Phys. Rev. B}, 84:113107, Sep 2011.

\bibitem{Kumar:2012}
Manohar Kumar, R{\'e}mi Avriller, Alfredo~Levy Yeyati, and Jan~M van
  Ruitenbeek.
\newblock Detection of vibration-mode scattering in electronic shot noise.
\newblock {\em Phys. Rev. Lett.}, 108(14):146602, 2012.

\bibitem{Frederiksen:2004}
Thomas Frederiksen, Mads Brandbyge, Nicol{\'a}s Lorente, and Antti-Pekka Jauho.
\newblock Inelastic scattering and local heating in atomic gold wires.
\newblock {\em Phys. Rev. Lett.}, 93(25):256601, 2004.

\bibitem{Wheeler:2013}
P.~J. Wheeler, Ruoyu Chen, and D.~Natelson.
\newblock Noise in electromigrated nanojunctions.
\newblock {\em Phys. Rev. B}, 87:155411, Apr 2013.

\bibitem{Muller2:1992}
CJ~Muller, JM~Van~Ruitenbeek, and LJ~De~Jongh.
\newblock Experimental observation of the transition from weak link to tunnel
  junction.
\newblock {\em Physica C: Superconductivity}, 191(3):485--504, 1992.

\bibitem{Muller:1992}
C.~J. Muller, J.~M. van Ruitenbeek, and L.~J. de~Jongh.
\newblock Conductance and supercurrent discontinuities in atomic-scale metallic
  constrictions of variable width.
\newblock {\em Phys. Rev. Lett.}, 69:140--143, Jul 1992.

\bibitem{Vrouwe:2005}
S.~A.~G. Vrouwe, E.~van~der Giessen, S.~J. van~der Molen, D.~Dulic, M.~L.
  Trouwborst, and B.~J. van Wees.
\newblock Mechanics of lithographically defined break junctions.
\newblock {\em Phys. Rev. B}, 71:035313, Jan 2005.

\bibitem{Wu:2007}
F~Wu, P~Queipo, A~Nasibulin, T~Tsuneta, TH~Wang, E~Kauppinen, and PJ~Hakonen.
\newblock Shot noise with interaction effects in single-walled carbon
  nanotubes.
\newblock {\em Phys. Rev. Lett.}, 99(15):156803, 2007.

\bibitem{Wheeler:2010}
PJ~Wheeler, JN~Russom, K~Evans, NS~King, and D~Natelson.
\newblock Shot noise suppression at room temperature in atomic-scale au
  junctions.
\newblock {\em Nano Lett.}, 10(4):1287--1292, 2010.

\bibitem{Chen:2014}
Ruoyu Chen, Patrick~J Wheeler, M~Di~Ventra, and D~Natelson.
\newblock Enhanced noise at high bias in atomic-scale au break junctions.
\newblock {\em Sci. Rep.}, 4:4221, 2014.

\bibitem{Spietz:2003}
Lafe Spietz, KW~Lehnert, I~Siddiqi, and RJ~Schoelkopf.
\newblock Primary electronic thermometry using the shot noise of a tunnel
  junction.
\newblock {\em Science}, 300(5627):1929--1932, 2003.

\bibitem{Adak:2015}
Olgun Adak, Ethan Rosenthal, Jeffrey Meisner, Erick~F Andrade, Abhay Pasupathy,
  Colin Nuckolls, Mark~S Hybertsen, and Latha Venkataraman.
\newblock Flicker noise as a probe of electronic interaction at metal-single
  molecule interfaces.
\newblock {\em Nano Lett.}, 2015.

\bibitem{Kumar:2013}
Manohar Kumar, Oren Tal, Roel~HM Smit, Alexander Smogunov, Erio Tosatti, and
  Jan~M van Ruitenbeek.
\newblock Shot noise and magnetism of pt atomic chains: Accumulation of points
  at the boundary.
\newblock {\em Phys. Rev. B}, 88(24):245431, 2013.

\bibitem{Vardimon:2015}
Ran Vardimon, Marina Klionsky, and Oren Tal.
\newblock Indication of complete spin filtering in atomic-scale nickel oxide.
\newblock {\em Nano Lett.}, 2015.

\bibitem{Armstrong:2010}
Jason~N Armstrong, RM~Schaub, Susan~Z Hua, and Harsh~Deep Chopra.
\newblock Channel saturation and conductance quantization in single-atom gold
  constrictions.
\newblock {\em Phys. Rev. B}, 82(19):195416, 2010.

\bibitem{Chen2:2014}
Ruoyu Chen, Manuel Matt, Fabian Pauly, Peter Nielaba, Juan~Carlos Cuevas, and
  Douglas Natelson.
\newblock Shot noise variation within ensembles of gold atomic break junctions
  at room temperature.
\newblock {\em Journal of Physics: Condensed Matter}, 26(47):474204, 2014.

\bibitem{Nielsen:2002}
SK~Nielsen, Mads Brandbyge, K~Hansen, Kurt Stokbro, JM~Van~Ruitenbeek, and
  Flemming Besenbacher.
\newblock Current-voltage curves of atomic-sized transition metal contacts: an
  explanation of why au is ohmic and pt is not.
\newblock {\em Phys. Rev. Lett.}, 89(6):066804, 2002.

\bibitem{Vardimon2:2015}
Ran Vardimon and Oren Tal.
\newblock Unpublished.

\end{thebibliography}

\end{document}